\documentclass[twocolumn,aps,pra,showpacs,preprintnumbers,10pt]{revtex4-1}
\usepackage{mathrsfs}
\usepackage{amsmath}
\usepackage{amsfonts}
\usepackage{amssymb}
\usepackage{amsthm}
\usepackage{graphicx}
\usepackage{color}
\usepackage{graphics}
\usepackage{bm}

\providecommand{\U}[1]{\protect\rule{.1in}{.1in}}
\begin{document}
\title{Thermal spin current and spin accumulation at ferromagnetic
insulator/nonmagnetic metal interface}
\author{Y.H. Shen}
\affiliation{Physics Department, The Hong Kong University of Science
and Technology, Clear Water Bay, Kowloon, Hong Kong}
\affiliation{HKUST Shenzhen Research Institute, Shenzhen 518057, China}
\author{X.S. Wang}
\email[Corresponding author:]{justicewxs@ust.hk}
\affiliation{Physics Department, The Hong Kong University of Science
and Technology, Clear Water Bay, Kowloon, Hong Kong}
\affiliation{HKUST Shenzhen Research Institute, Shenzhen 518057, China}
\author{X.R. Wang}
\email[Corresponding author:]{phxwan@ust.hk}
\affiliation{Physics Department, The Hong Kong University of Science
and Technology, Clear Water Bay, Kowloon, Hong Kong}
\affiliation{HKUST Shenzhen Research Institute, Shenzhen 518057, China}

\begin{abstract}
Spin current injection and spin accumulation near a ferromagnetic
insulator (FI)/nonmagnetic metal (NM) bilayer film under a thermal
gradient is investigated theoretically. Using the Fermi golden rule
and the Boltzmann equations, we find that FI and NM can exchange spins
via interfacial electron-magnon scattering because of the imbalance
between magnon emission and absorption caused by either non-equilibrium
distribution of magnons or non-equilibrium between magnons and electrons.
A temperature gradient in FI and/or a temperature difference across the
FI/NM interface generates a spin current which carries angular momenta
parallel to the magnetization of FI from the hotter side to the colder one.
Interestingly, the spin current induced by a temperature gradient in NM
is negligibly small due to the nonmagnetic nature of the non-equilibrium
electron distributions. The results agree well with all existing experiments.
\end{abstract}
\maketitle

\section{Introduction}
One of the important topics in spintronics is the spin current generation and
detection \cite{spintronics}. Compare with the electron spin current, the
magnon spin current has the advantage of lower energy consumption and longer
coherence time, especially in ferromagnetic insulators (FI) \cite{Kajiwara}.
Furthermore, magnons can be used to manipulate the motion of magnetic domain
walls \cite{magnon,Xiansi}. Recently, inter-conversion between electron
spin current and magnon spin current and various methods for magnon spin
current generation in FI were proposed, such as ferromagnetic resonance
(FMR) for coherent magnon spin current generation (known as spin pumping)
\cite{Kajiwara,Goennenwein,Hillebrands,Wumingzhong} and temperature gradient
for incoherent magnon spin current generation (known as spin Seebeck effect)
\cite{sse1,sse2,sse3,sse4,Goennenwein,phononsse}. The magnon spin current
can be detected by a nonmagnetic metal (NM) such as Pt or Pd with strong
spin-orbit couplings by which a spin current can be converted into an
electric current via inverse spin Hall effect (ISHE) \cite{ISHE,bauer}.
Since spin carriers in FI and NM are different (magnons in FI and electrons
in NM), the spin transport and spin current conversion between electrons
and magnons across the FI/NM interface becomes an interesting and important
issue for both the experiment interpretation and potential applications.

Different approaches have been used to investigate the spin transport in
FI/NM bilayer. The stochastic LLG equation coupled with ``spin mixing
conductance" concept \cite{YT,Goennenwein} describes successfully how a
spin current is pumped from FI into NM at FMR or under a temperature
gradient \cite{Xiaojiang,YT2}. However, the microscopic picture
of the spin pumping and spin mixing conductance was not given.
A quantum mechanical model based on interfacial $s-d$ coupling between
conducting electrons in NM and local magnetic moments in FI was also
proposed \cite{Maekawa,YT2012,SZhang,renjie} for spin Seebeck effect (SSE).
This model was originally designed for the transverse SSE \cite{sse2,sse3,
Maekawa}. In order to explain why spin current in NM changes direction in
the higher and the lower temperature sides of a sample, coupling of
phonons with spins and electrons is necessary \cite{Maekawa} if other
effect like the anomalous Nernst effect \cite{Chien} was not considered.
It is believed that a temperature gradient perpendicularly applied to
the interface (known as \textit{longitudinal} SSE \cite{sse3,sse4}) is a
clean configuration \cite{Chien} for SSE. In this paper, we investigate
the spin transport across FI/NM interface due to interfacial
electron-magnon interaction under a perpendicular temperature gradient.
Phonons do not dominate spin transport in this case, and are neglected.
We show that there is neither spin accumulation nor spin current at
thermal equilibrium, in consistent with the laws of thermodynamics.
Once there is a temperature gradient in the sample or a temperature
difference at the interface, spin accumulation occurs and a spin
current flows across the interface. Spins parallel to the magnetization
of FI flow from the hotter side to the colder one under a temperature
gradient in FI or under a temperature difference across the interface.
Surprisingly, a temperature gradient in NM cannot efficiently generate
a spin current because the spin currents from non-equilibrium spin-up
electrons and spin-down electrons cancel each other, resulting in a
negligible contribution. Our results are in good agreement with the
present experiments.

\begin{figure}
\begin{center}
\includegraphics[width=8.5cm]{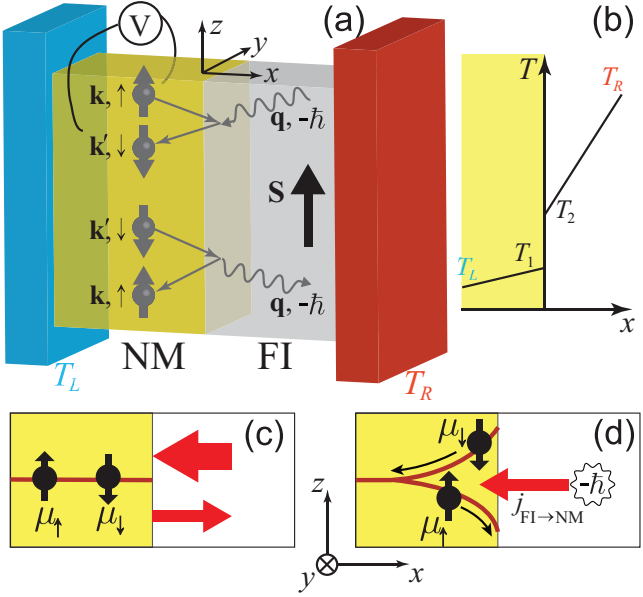}
\end{center}
\caption{(color online) (a) The setup of a FI/NM bilayer model
and possible electron-magnon scattering process at the interface:
a spin-up (spin-down) electron of wavevector $\mathbf{k}$
($\mathbf{k}^\prime$) becomes a spin-down (spin-up) electron of
wavevector $\mathbf{k}^\prime$ ($\mathbf{k}$) after absorbing
(emitting) a magnon of wavevector $\mathbf{q}$. NM (left) and FI
layers (right) are colored yellow and white. The blue and red blocks
denote respectively thermal reservoirs of temperature $T_L$ and $T_R$.
(b) Schematic diagram of the temperature profile when $T_R>T_L$.
(c) The instant when FI and NM are brought in contact. More magnons flow
to the left (thicker arrow) than those to the right (thinner arrow).
(d) At the steady state under a thermal gradient, spins flow across the
interface and spin accumulation occurs in NM near the  FI/NM interface.
Spin angular momentum alon the $-z$-direction flows from the hotter 
side to the colder side. 
}
\label{system}
\end{figure}

\section{Model and interfacial electron-magnon scattering}
Following the longitudinal SSE experiments \cite{sse3},
we consider a FI/NM bilayer model as shown in Fig. \ref{system}(a).
An NM layer is in contact with a FI layer, and two thermal reservoirs
of temperatures $T_L$ and $T_R$ are attached to the left side of NM and
and the right side of FI. The volume, thickness and lattice constant
of FI and NM layers are denoted by $V_i$, $d_i$ and $b_i$ ($i$=FI, NM).
The interface is in the $yz$-plane and its area is $A$.
Although most insulators used in SSE experiments are ferrimagnetic,
the energy of inter-sublattice excitations is too high to be excited
at low temperature \cite{Maekawa,Kaplan}. Only the acoustic spin waves
are relevant so that FIs are considered. A FI can be modeled by the
Heisenberg model of spin $S$ on cubic lattice. The electrons in NM are
modeled as a free-electron gas. Without losing generality, magnetization
of FI is in the $-z$-direction so that atomic spins $\mathbf{S}$
are in the $+z$-direction due to the negative gyromagnetic ratio.
The z-component of spins carried by spin-up (spin-down) electrons and
magnons are $\frac{\hbar}{2}\ (-\frac{\hbar}{2})$ and $-\hbar$
respectively in our model. The current density of spins along $-z$-direction 
in NM is $\mathbf{j}_s=\left(\frac{\hbar}{2e}\right)\mathbf{j}_\uparrow
-\left(\frac{\hbar}{2e}\right)\mathbf{j}_\downarrow$,
where $\mathbf{j}_{\uparrow\left(\downarrow\right)}$ is the electric
current density carried by spin-up (spin-down) electrons and
$e>0$ is the absolute value of the electron charge.

The interaction between electrons in NM and local magnetic moment is
modeled by an interfacial $s-d$ Hamiltonian \cite{grossobook,Goodings},
\begin{equation}
H=-\mathcal{J}_{sd}b^3_{\mathrm{FI}}\sum_n{\mathbf{s}\cdot\mathbf{S}_n}
\delta{(\mathbf{r}-\mathbf{R}_n)},
\label{H}
\end{equation}
where $\mathbf{s}$ is itinerant electron spin in NM and $\mathbf{S}_n$ is
local atomic spin at site $n$ of position $\mathbf{R}_n$ on the interface.
$\mathbf{s}$ and $\mathbf{S}_n$ are in the units of $\hbar$.
$\mathcal{J}_{sd}$ is the $s-d$ coupling strength and the summation is over
the atom sites on the interface. To calculate interfacial electron-magnon
scattering rate, we use the lowest-order Holstein-Primakoff transformation
\cite{H-P} $S_{n-}=\sqrt{2S}a_n^\dagger $ and $S_{n+}=\sqrt{2S}a_n $
($S_{n+}$ and $S_{n-}$ are ladder operators of $\mathbf{S}_n$ at site $n$
and $a^{\dagger}_n$, $a_n$ are the corresponding creation and annihilation
operators of magnons) so that magnon-magnon interactions are neglected.
The scattering due to the non-spin-flipping part of $H$ do not
contribute to spin current and spin accumulation, and is neglected.
In the momentum space, $H$ involving spin-flipping can be written as,
\begin{multline}
H^\prime=-\mathcal{J}_{sd}\frac{b^3_{\mathrm{FI}}N_\mathrm{IN}}{V_\mathrm{NM}}
\sqrt{\frac{S}{2N_\mathrm{FI}}}\times\\
\sum_{\mathbf{k},\mathbf{k^\prime},\mathbf{q}}(c^{\dagger}_{\mathbf{k}
\uparrow}c_{\mathbf{k}^\prime\downarrow}a^{\dagger}_{\mathbf{q}}
+c^{\dagger}_{\mathbf{k}^\prime\downarrow}c_{\mathbf{k}\uparrow}a_{\mathbf{q}}
)\delta_{\mathbf
{k}^\prime_{\parallel}-\mathbf{k}_{\parallel}=\mathbf{q}_{\parallel}},
\label{H2}
\end{multline}
where $c^{\dagger}_{\mathbf{k}\uparrow}$ ($c_{\mathbf{k}\uparrow}$)
and $c^{\dagger}_{\mathbf{k}\downarrow}$ ($c_{\mathbf{k}\downarrow}$)
are the creation (annihilation) operators of spin-up and spin-down
electrons of wavevector $\mathbf{k}$, respectively.
$a^{\dagger}_{\mathbf{q}}$ ($a_{\mathbf{q}}$) is the creation
(annihilation) operator of magnons of wavevector $\mathbf{q}$.
$N_\mathrm{FI}$ and $N_\mathrm{IN}$ are the numbers of atomic spins
in FI and at the interface, respectively. The first (second) term
describes an incident spin-down (spin-up) electron of wavevector
$\mathbf{k}'$ ($\mathbf{k}$) emitting (absorbing)
a magnon of wavevector $\mathbf{q}$ and becoming an outgoing spin-up
(spin-down) electron of wavevector $\mathbf{k}$ ($\mathbf{k}'$), as
illustrated in Fig. \ref{system}(a). This Hamiltonian preserves angular
momentum, and the momentum parallel to the interface is conserved.

Similar to the usual phonon-electron scattering calculation \cite{phonon}
by the Fermi golden rule, the magnon emission and absorption rates
between electron states $\mathbf{k}$ and $\mathbf{k}^\prime$ are
\begin{equation}
\begin{split}
t_\mathrm{em}&=S\frac{\pi}{\hbar}\frac{\mathcal{J}_{sd}^2}{V_\mathrm{NM}^2}
\frac{N_\mathrm{IN}^2}{N_\mathrm{FI}}[n(\mathbf{q})+1]\delta (E_{\mathbf{k}}
+\varepsilon_\mathbf{q}-E_{\mathbf{k^\prime}})\delta_{\mathbf{k}^\prime
_{\parallel}-\mathbf{k}_{\parallel}=\mathbf{q}_{\parallel}},\nonumber \\
t_\mathrm{ab}&=S\frac{\pi}{\hbar}\frac{\mathcal{J}_{sd}^2}{V_\mathrm{NM}^2}
\frac{N_\mathrm{IN}^2}{N_\mathrm{FI}} n(\mathbf{q})\delta(E_{\mathbf{k}}
+\varepsilon_\mathbf{q}-E_{\mathbf{k^\prime}})\delta_{\mathbf{k}^\prime
_{\parallel}-\mathbf{k}_{\parallel}=\mathbf{q}_{\parallel}},\nonumber \\
\end{split}
\end{equation}
where $n(\mathbf{q})$ is the number of magnons of wavevector $\mathbf{q}$,
$E_{\mathbf{k}}$ and $\varepsilon_\mathbf{q}$ are electron energy of
wavevector $\mathbf{k}$ and magnon energy of wavevector $\mathbf{q}$,
respectively. According to the physical picture illustrated in Fig.
\ref{system}(a), the perpendicular wavevector components should satisfy
$k'_x>0$, $k_x<0$, $q_x>0$ for magnon emission and $k_x>0$, $k'_x<0$,
$q_x<0$ for absorption. For simplicity, a quadratic dispersion is assumed
for electrons in NM, $E_\mathbf{k}=\frac{\hbar^2 |\mathbf{k}|^2}{2m}$.
The magnon spectrum is $\varepsilon_\mathbf{q}=J|\mathbf{q}|^2+D$
where $J$ is the ferromagnetic exchange coupling and $D$
is the gap due to magnetic anisotropy.

The net spin current density $j_\mathrm{FI\rightarrow NM}$ at the
interface is defined as the angular momentum parallel to the
magnetization of FI cross the interface per unit area and per unit time,
which is proportional to the difference of the absorbed magnon number
$N_\mathrm{ab}$ and the emitted one $N_\mathrm{em}$ per unit time,
\begin{equation}
j_\mathrm{FI\rightarrow NM}=\hbar\frac{N_\mathrm{ab}-N_\mathrm{em}}{A}.
\label{J_IN_1}
\end{equation}
By including the Pauli principle for electrons,
$N_\mathrm{ab}$ and $N_\mathrm{em}$ can be obtained from
$t_\mathrm{ab}$ and $t_\mathrm{em}$,
\begin{equation}
\begin{split}
N_\mathrm{em}&=\sum_{\mathbf{k},\mathbf{k^\prime},\mathbf{q}}
f_\downarrow(\mathbf{k'})[1-f_\uparrow(\mathbf{k})]t_\mathrm{em},\\
N_\mathrm{ab}&=\sum_{\mathbf{k},\mathbf{k^\prime},\mathbf{q}}
f_\uparrow(\mathbf{k})[1-f_\downarrow(\mathbf{k'})]t_\mathrm{ab},\\
\end{split}
\label{rates}
\end{equation}
where $f_s(\mathbf{k})$ is the electron distribution function of
wavevector $\mathbf{k}$ and spin $s=\uparrow,\ \downarrow$.
For a macroscopic system the summation can be converted into integration by
$\sum_{\mathbf{k},\mathbf{k^\prime},\mathbf{q}}\delta_{\mathbf{k}^\prime_
{\parallel}-\mathbf{k}_{\parallel}=\mathbf{q}_{\parallel}}\rightarrow
\frac{V_\mathrm{NM}}{(2\pi)^3}\frac{V_\mathrm{NM}}{(2\pi)^3}\frac{d_\mathrm
{FI}}{2\pi}\int\delta(\mathbf{k}_{\parallel}+\mathbf{q}_{\parallel}-
\mathbf{k}^\prime_{\parallel}){\rm{d}} \mathbf{k} {\rm{d}}\mathbf{k}' {\rm{d}} \mathbf{q}$.
The range of integration is $k_x>0$, $k'_x<0$, $q_x<0$ for magnon absorption
and $k'_x>0$, $k_x<0$, $q_x>0$ for emission. $\mathbf{q}$ is in the first
Brillioun Zone. To combine two integrals in Eq. \eqref{rates} together, we
change the dummy variables in $N_\mathrm{em}$ as $\mathbf{k}\to -\mathbf{k}$,
$\mathbf{k}^\prime \to -\mathbf{k}^\prime$ and $\mathbf{q}\to -\mathbf{q}$.
The spin current becomes
\begin{multline}
j_\mathrm{FI\rightarrow NM}=\hbar C\int_\mathrm{all} \{
f_\uparrow(\mathbf{k})[1-f_\downarrow(\mathbf{k'})]
n(\mathbf{q})\\
-f_\downarrow(\mathbf{-k'})[1-f_\uparrow(\mathbf{-k})][n(\mathbf{-q})+1]\}
\label{jtot}
\end{multline}
with $C=\frac{\pi}{\hbar}\frac{S\mathcal{J}_{sd}^2b_{\mathrm{FI}}^5}{(2\pi)^7}$.
Here $\int_\mathrm{all}=\int \delta(E_{\mathbf{k}}+\varepsilon_\mathbf{q}-
E_{\mathbf{k^\prime}})\delta(\mathbf
{k}_{\parallel}+\mathbf{q}_{\parallel}-\mathbf{k}^\prime_{\parallel})
\rm{d}  \mathbf{k} \rm{d} \mathbf{k}'\rm{d}  \mathbf{q}$ with $k_x>0$, 
$k'_x<0$, $q_x<0$ and $\mathbf{q}\in$Brillioun Zone.

\section{Spin transport at thermal equilibrium}\label{equil}

First we consider the case of the bilayer at thermal equilibrium ($T_L=T_R=T$).
The magnon number follows the Bose-Einstein distribution $n(\mathbf{q})
=n_0(\mathbf{q})=\frac{1}{e^{\beta \varepsilon_\mathbf{q}}-1}$ and the
electron distribution function is the Fermi-Dirac function
$f_s(\mathbf{k})=f_0(\mathbf{k})=\frac{1}{e^{\beta(E_{\mathbf{k}}-\mu_s)}
+1}$, where $s=\uparrow,\ \downarrow$, $\beta=(k_B T)^{-1}$ and $k_B$
is the Boltzmann constant. Because electrons are unpolarized in NM,
the chemical potentials of spin-up and spin-down electrons must be
the same, $\mu_\uparrow=\mu_\downarrow=\mu_0$, at the instant when FI
and NM are brought to contact. Due to the energy conservation
$E_\mathbf{k}+\varepsilon_\mathbf{q}=E_\mathbf{k'}$, we have
\begin{equation}
\begin{split}
f_0(\mathbf{k})&[1-f_0(\mathbf{k'})]n_0(\mathbf{q})\\
=&\frac{1}{e^{\beta(E_{\mathbf{k}}-\mu_0)}+1}\frac{e^{\beta
(E_{\mathbf{k'}}-\mu_0)} }{e^{\beta(E_{\mathbf{k'}}-\mu_0)}+1}
\frac{1}{e^{\beta \varepsilon_\mathbf{q}}-1}\\
=&\frac{1}{e^{\beta(E_{\mathbf{k}}-\mu_0)}+1}\frac{{e^{\beta (E_{\mathbf{k}}-\mu_0)}} e^{\beta \varepsilon_{\mathbf{q}}}}
{e^{\beta(E_{\mathbf{k'}}-\mu_0)}+1}\frac{1}{e^{\beta \varepsilon_\mathbf{q}}-1}\\
=&f_0(\mathbf{-k'})[1-f_0(\mathbf{-k})][n_0(\mathbf{-q})+1].
\end{split}
\label{detail_bal}
\end{equation}
Eq. \eqref{detail_bal} is the detailed balance between magnon absorption
and magnon emission at the thermal equilibrium. Using this
detailed balance result, Eq. \eqref{jtot} gives a vanishing spin current,
$j_\mathrm{FI\rightarrow NM}=0$, and no spin accumulation in this case.
In fact, no spin current and no spin accumulation at the thermal equilibrium
hold in general. Otherwise, a spin current would convert into charge current
via the ISHE effect. Thus, this device would generate electricity at the
thermal equilibrium! Since no external energy source exists in the set-up,
this assumption leads to a continuous extraction of electric energy from
a sole heat bath, a clear violation of the second law of thermodynamics.
Thus, our result must be model independent and true in general.
Also, if one regards spin accumulation as a proximity effect of a NM in
contact with a FI, this result says that the proximity effect does not exist
at the thermal equilibrium, very different from other types of proximity
effects such as a semiconductor carbon nanotube in contact with a metallic
carbon nanotube \cite{lujie}. There, the semiconductor carbon
nanotube becomes a weak metal at the thermal equilibrium.

\section{Spin transport at non-equilibrium }

When different temperatures $T_L$, $T_R$ are applied on the two sides of
the FI/NM bilayer as shown in Fig. 1(a), the system is at a non-equilibrium
state and thermal gradients will eventually established in both FI and NM.
Also, a temperature difference across the FI/NM interface may exist when
the thermal contact resistance is non-zero. The temperature profile can in
principle be obtained by solving the corresponding heat diffusion equations
with proper boundary conditions if the thermal conductivities and other
material parameters are known. Since the temperature profile is not the
subject of this work, we shall simply assume constant thermal conductivities
$\kappa_i$ of the materials ($i=\mathrm{FI,NM}$), and a constant
thermal contact resistance $R$ \cite{sse3,interresist,temperature}.
Thus, as shown in Fig. \ref{system}(b), a uniform temperature gradient of
$\alpha_\mathrm{NM}=(T_1-T_L)/d_\mathrm{NM}$ in NM, a uniform temperature
gradient of $\alpha_\mathrm{FI}=(T_R-T_2)/d_\mathrm{FI}$ in FI, and an
interfacial temperature difference $\Delta T\equiv T_2-T_1$
across the interface are established at the steady state.
$\alpha_\mathrm{NM}$, $\alpha_\mathrm{FI}$ and $\Delta T$ satisfy
\begin{equation}
\begin{split}
&\alpha_\mathrm{NM}d_\mathrm{NM}+\Delta T+\alpha_\mathrm{FI}d_\mathrm{FI}
=T_R-T_L,\nonumber \\
&\alpha_\mathrm{NM}\kappa_\mathrm{NM}=\frac{\Delta T}{R}=
\alpha_\mathrm{FI}\kappa_\mathrm{FI}. \nonumber
\end{split}
\end{equation}
$\alpha_\mathrm{NM}$ will induce a non-equilibrium distribution
of electrons, while $\alpha_\mathrm{FI}$ will induce a
non-equilibrium distribution of magnons. $\Delta T$ will
break the detailed balance between the magnon absorption
and emission as shown in section III.
Since the magnon emission and absorption are no longer balanced,
a net spin current across the interface shall appear.

On the other hand, due to the spin conserved $s-d$ interaction
at the interface, each absorbed magnon results in an electron to
flip from spin-up to spin-down state, and each emitted magnon
causes an electron to flip from spin-down to spin-up state.
Thus, if there are more absorbed magnons than emitted ones
($j_{\rm{FI\rightarrow NM}}>0$ according to Eq. \eqref{J_IN_1}), the
number of spin-down electrons would be larger than that of spin-up
electrons, and chemical potential of spin-up and spin-down electrons
would no longer be the same and $\mu_{\downarrow}>\mu_{\uparrow}$.
Similarly, when $j_{\rm{FI\rightarrow NM}}<0$,
we have $\mu_{\uparrow}>\mu_{\downarrow}$.
The electron spin accumulation near the interface causes a
spin current $\mathbf{j}_{s}$ in NM due to spin diffusion.
This spin current should be continuum at the interface. 
Thus we have
\begin{equation}
-j_{\rm{FI\rightarrow NM}}=j_{s}(0)
=\left(\frac{\hbar}{2e}\right)j_{\uparrow}(0)
+\left(-\frac{\hbar}{2e}\right)j_{\downarrow}(0).
\label{b.c.}
\end{equation}
Both $j_{\uparrow (\downarrow)}$ and $j_{\rm{FI\rightarrow NM}}$ can be
determined by the distribution functions of electrons and magnons, which
will be studied by solving the Boltzmann equations in the next subsection.

\subsection{Distribution functions under given temperature profile}

When the system is not far from the equilibrium, $n(\mathbf{q})$ and
$f_s(\mathbf{k})$ ($s=\uparrow,\downarrow$) are govern
by the Boltzmann equations with the relaxation time approximation.
For magnons, the distribution function $n(\mathbf{q})$ under the thermal
gradient $\alpha_{\rm{FI}}$ can be obtained by solving following
Boltzmann equation
\begin{equation}
\mathbf{v}(\mathbf{q})\cdot\nabla n(\mathbf{q})
=-\frac{n_{1}(\mathbf{q})}{\tau},
\label{B.E._m}
\end{equation}
where $\mathbf{v}(\mathbf{q})=\frac{1}{\hbar}\nabla_\mathbf{q}\varepsilon_q$
is the group velocity of magnons with wavevector $\mathbf{q}$,
$\tau$ is the average relaxation time of magnons,
and $n_1(\mathbf{q})=n(\mathbf{q})-n_0(\mathbf{q})$ where
$n_0(\mathbf{q})=\frac{1}{e^{\beta\varepsilon_q}-1}$
is the Bose-Einstein distribution with the local temperature.
To the first order in $\alpha_{\rm FI}$, we can replace $n(\mathbf{q})$ by
$n_{0}(\mathbf{q})$ in Eq. \eqref{B.E._m}, and obtain
\begin{equation}
n(\mathbf{q})=n_{0}(\mathbf{q})-\tau v_{x}(\mathbf{q})\alpha_{\rm FI}
\frac{\partial n_{0}(\mathbf{q})}{\partial T},
\label{n}
\end{equation}
where $\frac{\partial n_0}{\partial T}=\frac{\beta\varepsilon_q}{T}n_0(n_0+1)$.
Obviously, we have $n_{1}(-\mathbf{q})=-n_{1}(\mathbf{q})$.

For electrons, the non-equilibrium distribution is not only affected
by the temperature gradient $\alpha_{\rm{NM}}$, but also by the spin
accumulation near the interface, as shown in Fig. \ref{system}(d).
To take this spin accumulation into consideration,
we need to solve the Boltzmann equation about $f_s(\mathbf{k},\mathbf{r})$
including the spin-flip process \cite{Fert, SZhang2000, SZhang}:
\begin{equation}
\mathbf{v}\cdot\nabla f_s
+\left(-\frac{e\bm{\mathcal{E}}}{\hbar}\right)\cdot\nabla_{\mathbf{k}}f_s
=-\frac{f_s-f_{0,s}}{\tau_{c}}-\frac{f_{0,s}-f_{0,-s}}{\tau_{sf}},
\label{B.E._f}
\end{equation}
where $\mathbf{v}=\frac{\hbar\mathbf{k}}{m}$,
$f_{0,s}=[e^{\beta(E-\mu_s)}+1]^{-1}$
is the equilibrium distribution function with local temperature and
local electrochemical potential $\mu_s(x)$ for spin $s$.
The relaxation times $\tau_{c}$ and $\tau_{sf}$ describe respectively
the momentum-energy relaxation and spin relaxation of electrons.
$\bm{\mathcal{E}}=-\nabla\phi$ is the electric field in NM, and
$E=\frac{\hbar^{2}k^{2}}{2m}-e\phi$ is the electron energy.
To solve Eq. \eqref{B.E._f} in linear response regime ($f_{1,s}$
linear in the temperature gradient and electric field), we can replace
$f_s$ by $f_{0,s}$ in the left-hand side of Eq. \eqref{B.E._f}.
Thus, we obtain
\begin{multline}
\mathbf{v}\cdot\left[\frac{-\nabla T}{T}\left(E-\mu_s\right)-\nabla\mu_s\right]\frac{\partial f_{0,s}}{\partial E}\\
=-\frac{f_s-f_{0,s}}{\tau_{c}}-\frac{f_{0,s}-f_{0,-s}}{\tau_{sf}}.
\label{B.E._f_1}
\end{multline}
Normally, the deviation of the local electrochemical potential
$\delta\mu_s=\mu_s-\mu_{e}$ from the electrochemical potential without spin
accumulation ($\mu_{e}$) is small. Since the change of the density of state
near the Fermi surface is small, it is common to use the approximation of
$\delta\mu_\uparrow=-\delta\mu_\downarrow$ \cite{Mosendz,Saslow,SZhang}.
After expanding $\mu_s$ and $f_{0,s}$ at $\mu_{e}$ and keeping the linear
terms in Eq. \eqref{B.E._f_1}, we have
\begin{equation}
f_s(\mathbf{k})=f_{0}(\mathbf{k})-\left(1-2\frac{\tau_{c}}{\tau_{sf}}\right)
\delta\mu_s\frac{\partial f_{0}(\mathbf{k})}{\partial E}+g_s(\mathbf{k}),
\label{f}
\end{equation}
where
\begin{equation}
g_s(\mathbf{k})=\tau_{c}v_{x}(\mathbf{k})
\left[\frac{\alpha_{\mathrm{NM}}}{T}\left(E_{k}-\mu_{e}\right)
+\frac{\mathrm{d}\mu_{e}}{\mathrm{d}x}+\frac{\mathrm{d}\delta\mu_s}{\mathrm{d}x}\right]
\frac{\partial f_{0}(\mathbf{k})}{\partial E},\nonumber
\end{equation}
and $f_0=[e^{\beta\left(E-\mu_e\right)}+1]^{-1}$ is the Fermi-Dirac
distribution function without spin accumulation. Because $\tau_c\ll\tau_{sf}$
in most cases,  we can discard $2\frac{\tau_c}{\tau_{sf}}$ in Eq. \eqref{f}.
Obviously, we have $g_s(-\mathbf{k})=-g_s(\mathbf{k})$ and
$f_{1,s}(\mathbf{k})=f_s(\mathbf{k})-f_0(\mathbf{k})=-\delta\mu_s
\frac{\partial f_0(\mathbf{k})}{\partial E}+g_s(\mathbf{k})$.

Since $\frac{\mathrm{d}\mu_e}{\mathrm{d}x}$ and $\frac{\mathrm{d}\delta\mu_\uparrow}
{\mathrm{d}x}=-\frac{\mathrm{d}\delta\mu_\downarrow}{\mathrm{d}x}$ are still unknown,
we need to consider the charge/spin transport in NM.
In NM where $(f_\uparrow+f_\downarrow)= 2f_0+ 2\tau_c v_x[\frac{\alpha_{
\rm NM}}{T}(E-\mu_e)+\frac{{\rm d}\mu_e}{{\rm d}x}]\frac{\partial
f_0}{\partial E}$, the electric current
\begin{equation}
j=\frac{(-e)}{V_{{\rm NM}}}\sum_{\mathbf{k}}2\tau_{c}v_{x}^{2}
[\frac{\alpha_{{\rm NM}}}{T}(E-\mu_{e})+\frac{{\rm d}\mu_{e}}{{\rm d}x}]
\frac{\partial f_{0}}{\partial E}
\label{current}
\end{equation}
is not affected by the spin accumulation, and the spin current
\begin{equation}
j_s=
-\frac{\hbar}{2V_{{\rm NM}}}\sum_{\mathbf{k}}v_{x}\left(f_{\uparrow}-f_{\downarrow}\right)
=\frac{\hbar\sigma}{2e^{2}}\frac{{\rm d}\delta\mu_{\uparrow}}{{\rm d}x}
\label{spincurrent}
\end{equation}
depends on the spin accumulation.
$\sigma=\frac{ne^2\tau_c}{m}$ is the conductivity of the metal, $n$ is the
electron density in the NM.

The distribution of $\delta\mu_\uparrow$ inside NM can be determined by
the diffusion equation \cite{Fert,Mosendz,Saslow,SZhang2000,SZhang}:
\begin{equation}
\frac{\mathrm{d}^2\delta\mu_\uparrow}{\mathrm{d}x^2}=
\frac{\delta\mu_\uparrow}{l_{sd}^2},\nonumber
\end{equation}
where $l_{sd}$ is the spin diffusion length. For $d_{\mathrm{NM}}\gg l_{sd}$,
$\delta\mu_{\uparrow}(x)=\delta\mu_{\uparrow}(0)\exp(x/l_{sd})$, and
\begin{equation}
\frac{\mathrm{d}\delta\mu_\uparrow}{\mathrm{d}x}=\frac{\delta\mu_\uparrow}{l_{sd}}.
\end{equation}

$\frac{{\rm d}\mu_e}{{\rm d}x}$ can be determined from the fact that there
is no electric current in an open circuit, Eq. \eqref{current} gives
\begin{equation}
\frac{{\rm d}\mu_e}{{\rm d}x}=-\frac{\alpha_{\rm NM}\int
v^2\left(E-\mu_e\right)\frac{\partial f_0}{\partial E}
{\rm d}^3\mathbf{k}}{T\int v^2 \frac{\partial f_0}{\partial E}
{\rm d}^3\mathbf{k}}\approx-\frac{\alpha_{\rm NM}}{2T}
\frac{\left(\pi k_{{\rm B}}T\right)^{2}}{\mu_{e}}.
\label{seebeck}
\end{equation}
This is the conventional Seebeck effect \cite{thermobook}.

To fully determine $f_s(\mathbf{k})$, one still needs to find out
$\delta\mu_{\uparrow}(0)$ in terms of known model parameters.
The right hand side of Eq. \eqref{b.c.} is linear in
$\delta\mu_\uparrow(0)$ after using expression found early for $j_s$ and
$\frac{\mathrm{d}\delta\mu_\uparrow}{\mathrm{d}x}=\frac{\delta\mu_\uparrow}{l_{sd}}$.
$j_{\rm{FI\rightarrow NM}}$ in Eq. \eqref{b.c.} can also be expressed by
$n(\mathbf{q})$ and $f_s(\mathbf{k})$ as given by Eq. \eqref{jtot}.
Thus Eq. \eqref{b.c.} would be an equation about $\delta\mu_{\uparrow}(0)$.
Then we can obtain the spin accumulation and spin current across the
interface as shown below.

\subsection{Spin current in linear response regime}

In the last subsection, we obtained $f_s(\mathbf{k})=f_{0}(\mathbf{k})+
f_{1,s}(\mathbf{k})$ and $n(\mathbf{q})=n_0(\mathbf{q})+n_1
(\mathbf{q})$, where $f_{1,s}$ and $n_1$, linear in thermal gradient,
is much smaller than their equilibrium values.
Substitute them into Eq. \eqref{jtot} and keep only the terms up to
linear orders in $f_{1,s}$ and $n_{1}$, the spin current can be
decomposed into three terms:
\begin{equation}
j_{\rm{FI\rightarrow NM}}=j_{d}+j_{m}+j_{e},
\end{equation}
where
\begin{multline}
j_d=\hbar C\int_\mathrm{all} \{ f_0(\mathbf{k})[1-f_0(\mathbf{k'})]
n_0(\mathbf{q})\\
-f_0(\mathbf{k'})[1-f_0(\mathbf{k})][n_0(\mathbf{q})+1]\},
\label{jd}
\end{multline}
\begin{multline}
j_m=\hbar C
\int_\mathrm{all} \{ f_0(\mathbf{k})[1-f_0(\mathbf{k'})]
n_1(\mathbf{q})
\\-f_0(\mathbf{k'})[1-f_0(\mathbf{k})]n_1(\mathbf{-q})\} ,
\label{jm}
\end{multline}
\begin{multline}
j_e=\hbar C\int_\mathrm{all} \{f_{1,\uparrow}(\mathbf{k})[1-f_0(\mathbf{k'})]
-f_0(\mathbf{k})f_{1,\downarrow}(\mathbf{k'})\}n_0(\mathbf{q})-\{ \\
f_{1,\downarrow}(\mathbf{-k'})[1-f_0(\mathbf{k})]
-f_0(\mathbf{k'})f_{1,\uparrow}(\mathbf{-k})\}[n_0(\mathbf{q})+1],
\label{je}
\end{multline}
where $f_0=\frac{1}{e^{\beta_1(E-\mu_e)}+1}$
and $n_0=\frac{1}{e^{\beta_2\varepsilon}-1}$ with
$\beta_1=\frac{1}{k_{\rm{B}}T_1}$,
$\beta_2=\frac{1}{k_{\rm{B}}T_2}$.
$T_1$ and $T_2$ are the temperatures of NM and FI at the FI/NM interface.

Since Eq. \eqref{detail_bal} is no longer valid if $\beta_1\neq\beta_2$, $j_d$
is not zero in this case. $j_m$ is mainly due to the deviation of magnons
from their equilibrium distribution. If there is no temperature gradient
$\alpha_{\rm{FI}},$  Eq. \eqref{n} says $n_1=0$, and then $j_m$ vanishes.
$j_e$ comes from the deviation of electrons from their equilibrium distribution.
According to Eq. \eqref{f}, the deviation is caused by both temperature
gradient $\alpha_{\rm{NM}}$ as well as the spin accumulation
$\delta\mu\uparrow$ originated from the spin injection across the interface.
Even if $\alpha_{\rm{NM}}=0$, $f_{1,s}$ still exists as long as there is
a nonzero spin current, for example from $\Delta T=T_2-T_1\neq0$ or
$\alpha_{\rm{FI}}\neq0$. Below, we will study $j_d$, $j_m$, $j_e$ separately,
and by applying the boundary condition given in Eq. \eqref{b.c.}, and find
out the relationship between spin current across the interface, spin
accumulation at steady state and $\Delta T$, $\alpha_{\rm{FI}}$,
$\alpha_{\rm{NM}}$. To simplify the presentation, we introduce two notations
\begin{equation}
\begin{split}
L_1&=f_0(\mathbf{k})[1-f_0(\mathbf{k'})]n_0(\mathbf{q}),\\
L_2&=f_0(\mathbf{k'})[1-f_0(\mathbf{k})][n_0(\mathbf{q})+1].
 \end{split}
 \label{L2}
 \end{equation}

Then, $j_d=\hbar C\int_\mathrm{all}(L_1-L_2)$, where
$L_1/L_2=e^{(\beta_1-\beta_2)\varepsilon_{q}}$.
When $T_2>T_1$, the magnons have higher temperature, we have $\beta_1>\beta_2$,
$L_1>L_2$, and $j_d>0$.
The spin current induced by the temperature difference flows from FI to NM.
The spin current reverses its direction when $T_1>T_2$.
In general, temperature difference at the interface generates a spin flow
from the hotter side to the colder side.
When $\Delta T\ll T_1,T_2$, we can expand $L_1/L_2\approx1+(\beta_1-\beta_2
)\varepsilon_{q}=1+\Delta T\frac{\varepsilon_{q}}{k_{\rm{B}}T_{1}T_{2}}$,
then $j_d=\mathcal{K}_{1} \Delta T$ is proportional to $\Delta T$, and
coefficient $\mathcal{K}_1$ is
\begin{equation}
\mathcal{K}_1
=\hbar C\int_\mathrm{all}\frac{\varepsilon_qL_2}{k_{\rm B}T_1T_2}.
\label{k1}
\end{equation}

To evaluate $j_m$, we substitute Eq. \eqref{n} into Eq. \eqref{jm}
and, noting that $n_1(-\mathbf{q})=-n_1(\mathbf{q})$, we obtain
\begin{multline}
j_m=\alpha_{\rm{FI}}\hbar C\int_{\mathrm{all}}\Big\{ \frac{-\tau
v_x(\mathbf{q})\varepsilon_q}{k_{\rm{B}}T_2^2}L_2\\
\left[(2n_0+1)+\Delta T\frac{\varepsilon_q}{k_{\rm B}T_1T_2}(n_0+1)\right]\Big\}.
 \label{jm1}
 \end{multline}
For $\Delta T\ll T_{1},T_{2}$, then
$L_1/L_2\approx1+\Delta T\frac{\varepsilon_q}{k_{\rm{B}}T_1T_2}$.
For the linear response of the spin current to $\Delta T$, $\alpha_{\rm{FI}}$
and $\alpha_{\rm{NM}}$, we can drop the last term in the bracket in
Eq. \eqref{jm1} that would result in a higher order contribution,
proportional to $\Delta T\cdot\alpha_{\rm{FI}}$.
Note that the integration range includes only $q_{x}<0$,
thus $-v_{x}(\mathbf{q})=-\frac{2J}{\hbar}q_{x}>0$,
$j_m=\mathcal{K}_{2}\alpha_{{\rm FI}}$ is proportional to $\alpha_{\rm{FI}}$,
and coefficient $\mathcal{K}_{2}$ is
\begin{equation}
\mathcal{K}_{2}=\hbar C\int_\mathrm{all}\frac{-\tau v_{x}\left(\mathbf{q}
\right)\varepsilon_{q}}{k_{{\rm B}}T_{2}^{2}}L_{2}\left(2n_{0}+1\right),
\label{k2}
\end{equation}
which is positive.
When $\alpha_{\rm{FI}}>0$ ($T_L<T_R$), the spin current flows from FI to NM,
and reverses its direction when $\alpha_{\rm{FI}}<0$.
In general, the spin current caused by temperature gradient in FI flows
from hotter side to colder side.

In order to compute $j_e$ and because $f_{1,s}$ contains many terms,
we decompose $j_e$ into $j_{e,i}$, due to the isotropic part
$-\delta\mu_s\frac{\partial f_{0}(\mathbf{k})}{\partial E}$ of $f_{1,s}$,
and $j_{e,a}$, due to the anisotropic part $g_s(\mathbf{k})$ of $f_{1,s}$,
\begin{multline}
j_{e,i}=\delta\mu_{\uparrow}\left(0\right)\hbar C\int_{\mathrm{all}}\beta_{1}L_{2}\\
\left\{ \left(\frac{L_{1}}{L_{2}}+1\right)+\left(\frac{L_{1}}{L_{2}}-1\right)\left[f_{0}\left(\mathbf{k}'\right)-f_{0}\left(\mathbf{k}'\right)\right]\right\}
 \label{jei}
 \end{multline}
\begin{equation}
\begin{split}
j_{e,a}=\hbar C\int_{\mathrm{all}}&\left[a_{\uparrow}\left(\mathbf{k}\right)\right]\left(-\beta_{1}L_{2}\right)\\
&\left\{ \left(\frac{L_{1}}{L_{2}}-1\right)\left[1-f_{0}\left(\mathbf{k}\right)\right]+\left[1-2f_{0}\left(\mathbf{k}\right)\right]\right\} \\
+&\left[a_{\downarrow}\left(\mathbf{k}'\right)\right]\left(-\beta_{1}L_{2}\right)\\
&\left\{ \left[1-2f_{0}\left(\mathbf{k}'\right)\right]-f_{0}\left(\mathbf{k}'\right)\left(\frac{L_{1}}{L_{2}}-1\right)\right\},
 \end{split}
 \label{jea}
 \end{equation}
 where
 \begin{equation}
 \begin{split}
 a_{\uparrow}\left(\mathbf{k}\right)=&\tau_{c}v_{x}\left(\mathbf{k}\right)
 \left[\frac{\alpha_{{\rm NM}}}{T_{1}}\left(E_{k}-\mu_{e}\right)
 +\frac{{\rm d}\mu_{e}}{{\rm d}x}+\frac{{\rm d}\delta\mu_{\uparrow}}{{\rm d}x}\right],\\
 a_{\downarrow}\left(\mathbf{k}'\right)=&\tau_{c}v_{x}\left(\mathbf{k}'\right)
 \left[\frac{\alpha_{{\rm NM}}}{T_{1}}\left(E_{k'}-\mu_{e}\right)
 +\frac{{\rm d}\mu_{e}}{{\rm d}x}-\frac{{\rm d}\delta\mu_{\uparrow}}{{\rm d}x}\right].
 \end{split}
 \end{equation}
Since $\delta\mu_{\uparrow}(0)$ is always small, we keeps only linear terms
so that all terms with $\left(\frac{L_{1}}{L_{2}}-1\right)$ in Eq. (26,27)
are neglected.
Then we have $j_{e,i}\approx\mathcal{K}_{3}\delta\mu_{\uparrow}\left(0\right)$
and $j_{e,a}\approx \mathcal{K}_4\alpha_{{\rm NM}}+\mathcal{K}_5\frac{{\rm d}
\mu_{e}}{{\rm d}x}+\mathcal{K}_6\frac{{\rm d}\delta\mu_{\uparrow}}{{\rm d}x}$
with coefficients $\mathcal{K}_i$ ($i=3,4,5,6$) being
\begin{equation}
\begin{split}
\mathcal{K}_3=& \hbar C\int_\mathrm{all}\left(\frac{2L_2}{k_{\rm B}T_1}\right),\\
\mathcal{K}_4=&\hbar C\int_\mathrm{all} \frac{-\tau_cL_2} {k_{\rm B}T_1^2} \{
v_x\left(\mathbf{k}\right) \left(E_{k}-\mu_{e}\right)\left[1-2f_0
\left(\mathbf{k}\right)\right]\\
&+v_x(\mathbf{k}') (E_{k'}-\mu_e)\left[1-2f_0(\mathbf{k}')\right]\},\\
\mathcal{K}_5=&\hbar C\int_\mathrm{all}\frac{-\tau_cL_2}{k_{\rm B}T_1}
\{ v_x(\mathbf{k})\left[1-2f_0(\mathbf{k})\right]+ \\
&v_x(\mathbf{k}')\left[1-2f_0(\mathbf{k}')\right] \} ,\\
\mathcal{K}_6=&\hbar C\int_\mathrm{all}\frac{-\tau_cL_2}{k_{\rm B}T_1}
\{ v_x(\mathbf{k})\left[1-2f_0(\mathbf{k})\right]-\\
&v_x(\mathbf{k}')\left[1-2f_0(\mathbf{k}')\right] \} .
\end{split}
\label{k3456}
\end{equation}
Note that $\mathcal{K}_{4},\ \mathcal{K}_{5},\ \mathcal{K}_{6}$ contain
a factor $\left[1-2f_{0}\left(\mathbf{k}\right)\right]$ in the integrals
and since only electrons near the Fermi surface participant in scatterings,
$\left[1-2f_{0}\left(\mathbf{k}\right)\right]L_{2}$ is always small.
For $\mathcal{K}_{5}$ and $\mathcal{K}_{6}$, the factor
$\left(1-2f_{0}\right)\approx\frac{1}{2}\beta_{1}\left(E-\mu_{e}\right)$
which change its sign at the Fermi surface, and the contribution from the
electrons above and below the Fermi surface almost cancel each other.
For $\mathcal{K}_{4}$, though $\left(E-\mu_{e}\right)\left(1-2f_{0}\right)>0$,
but noting that $k_{x}>0$ and $k_{x}'<0$, the two parts of $\mathcal{K}_{4}$
have different sign, and their magnitudes are almost the same.
According to the previous section, we have $\frac{{\rm d}\mu_e}{{\rm d}x}
=-\frac{\alpha_{NM}}{2T}\frac{\left(\pi k_{B}T\right)^{2}}{\mu_{e}}$ and
$\frac{{\rm d}\delta\mu_\uparrow}{{\rm d}x}=\frac{\delta\mu_\uparrow}{l_{sd}}$.
The non-equilibrium distribution of electrons would induce a spin
current as
\begin{eqnarray}
j_e=j_{e,i}+j_{e,a} & = & \mbox{\ensuremath{\mathcal{K}}}_3'\delta
\mu_\uparrow\left(0\right)+\mathcal{K}_4'\alpha_{NM}\
\end{eqnarray}
where $\mathcal{K}_3'=\mathcal{K}_{3}+\frac{1}{l_{sd}}\mathcal{K}_{6}$
and $\mathcal{K}_{4}'=\mathcal{K}_{4}-\frac{T\left(\pi k_{B}\right)^{2}}{2\mu_{e}}\mathcal{K}_{5}$.
Numerical results in the next section shows that $\mathcal{K}_{4}'$
is much smaller than $\mathcal{K}_{2}$ and almost zero.

\subsection{Spin current injection and spin accumulation at steady state}
Substitute results of $j_{\rm FI\rightarrow NM}$ obtained in the previous
subsection and Eq. \eqref{spincurrent} into Eq. \eqref{b.c.}, we have
\begin{eqnarray}
& \mathcal{K}_1& \Delta T+\mathcal{K}_2\alpha_{\rm FI}+
\mathcal{K}_4'\alpha_{\rm NM}+  \nonumber \\
& & \left( \mathcal{K}_3'+\frac{\hbar}{2e^2}\frac{\sigma}
{l_{sd}}\right)\delta\mu_{\uparrow}(0)=0,
\end{eqnarray}
and, thus spin accumulation at FI/NM interface is
\begin{eqnarray}
\delta\mu_{\uparrow}(0) & = & -\frac{2e^2l_{sd}\mathcal{K}_1}{2e^2
l_{sd}\mathcal{K}_3'+\hbar\sigma}\Delta T -\frac{2e^2l_{sd}\mathcal
{K}_2}{2e^2l_{sd}\mathcal{K}_3'+\hbar\sigma}\alpha_{\rm FI} \nonumber\\
&  & -\frac{2e^2l_{sd}\mathcal{K}_4'}{2e^2l_{sd}\mathcal{K}_3'+
\hbar\sigma}\alpha_{\rm NM}.
\end{eqnarray}
Then the total spin current across the interface is
\begin{eqnarray}
j_{\rm FI\rightarrow NM} & = & \frac{\hbar\sigma\mathcal{K}_1}{\hbar
\sigma+2e^2l_{sd}\mathcal{K}_3'}\Delta T+\frac{\hbar\sigma\mathcal{K}_2}
{\hbar\sigma+2e^2l_{sd}\mathcal{K}_3'}\alpha_{{\rm FI}} \nonumber \\
&  & +\frac{\hbar\sigma\mathcal{K}_4'}{\hbar\sigma+2e^2l_{sd}
\mathcal{K}_3'}\alpha_{\rm NM},
\label{total-spincurrent}
\end{eqnarray}
that flows from hotter side to colder side.
$\mathcal{K}_1,\mathcal{K}_2,\mathcal{K}_3',\mathcal{K}_4'$
are functions of $T_{1}$ and $T_{2}$, which can be determined by
Eqs. (22) , (24) and (28). $T_{1}$ and $T_{2}$ are determined by
model parameters, as shown in Section IV,
\begin{equation}
\begin{split}
T_1=&\frac{T_L(R\kappa_{\rm FI}\kappa_{\rm NM}+d_{\rm FI}\kappa_{\rm NM})
+T_Rd_{\rm NM}\kappa_{\rm FI}} {R\kappa_{\rm FI}\kappa_{\rm NM}+
d_{\rm NM}\kappa_{\rm FI}+d_{\rm FI}\kappa_{\rm NM}}, \\
T_2=&\frac{T_Ld_{\rm FI}\kappa_{\rm NM}+T_R(R\kappa_{\rm FI}\kappa_{\rm NM}
+d_{\rm NM}\kappa_{\rm FI})} {R\kappa_{\rm FI}\kappa_{\rm NM}+
d_{\rm NM}\kappa_{\rm FI}+d_{\rm FI}\kappa_{\rm NM}}.
\end{split}
\end{equation}

\section{Numerical Results and Discussion}

\begin{figure}[!htb]
\begin{center}
\includegraphics[width=8.5cm]{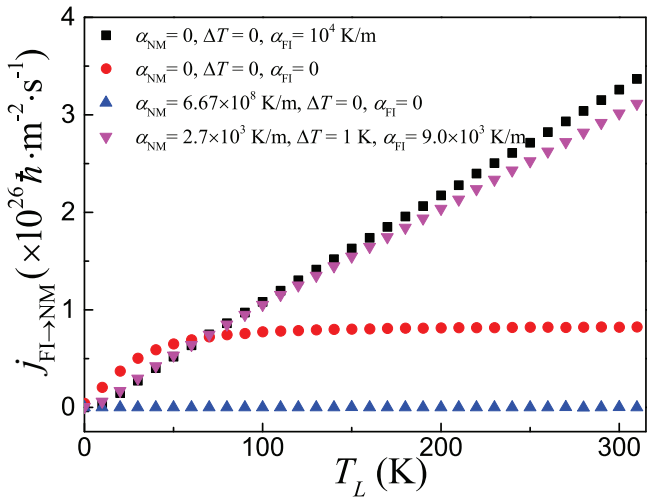}
\end{center}
\caption{(color online) Spin current $j_\mathrm{FI\rightarrow NM}$
as a function of $T_L$ when $T_R=T_L+10$ K
for various sets of $\alpha_\mathrm{NM}, \alpha_\mathrm{FI}$ and
$\Delta T$: $\alpha_\mathrm{NM}=0$, $\alpha_\mathrm{FI}=(T_R-
T_L)/d_\mathrm{FI}=10^4$ K/m, and $\Delta T=0$ (black squares);
$\alpha_\mathrm{NM}=\alpha_\mathrm{FI}=0$, and $\Delta T=10$ K
(red circles); $\alpha_\mathrm{NM}=(T_R-T_L)/
d_\mathrm{NM}=6.67\times 10^8$ K/m, $\alpha_\mathrm{FI}=0$, and
$\Delta T=0$ (blue up triangles); $\alpha_\mathrm{NM}=2.7\times
10^3$ K/m, $\alpha_\mathrm{FI}=9.0\times 10^3$ K/m, and
$\Delta T=1$ K (blue up triangles).}
\label{result}
\end{figure}

To have a better idea about the magnitude of the spin current and spin
accumulation generated by a thermal gradient, we numerically compute
the total spin current $j_\mathrm{FI\rightarrow NM}$ given by Eq.
\eqref{total-spincurrent} with realistic model parameters of YIG:
$S=23.6$, $J=1.9\times10^{-40}$ J$\cdot \mathrm{m}^2$, $D=1.8\times10^
{-24}$ J, $\tau=10^{-7}$ s \cite{SZhang,BEC}, $b_\mathrm{FI}=1.2$ nm;
and Pt: $\sigma=9.4\times10^6$ $\mathrm{m^{-1}\Omega^{-1}}$, $l_{sd}=1
.5$ nm, $\mu_0=9.74$ eV. $\mathcal{J}_{sd}=1$ meV, $d_\mathrm{FI}=1$ mm,
and $d_\mathrm{NM}=15$ nm \cite{sse3} are used.
The temperature difference between two thermal reservoirs is set to
$T_R-T_L=10$ K. In order to know which thermal source is
more effective in spin current generation, we first examine the cases when
all 10 K temperature difference is on FI, NM or at the FI/NM interface.
The results are shown in Fig. \ref{result} for $\alpha_\mathrm{NM}=0$,
$\alpha_\mathrm{FI}=(T_R-T_L)/d_\mathrm{FI}=10^4$ K/m,
and $\Delta T=0$ (black squares); $\alpha_\mathrm{NM}=\alpha_\mathrm{FI}=0$,
and $\Delta T=T_R-T_L=10$ K (red circles);
$\alpha_\mathrm{NM}=(T_R-T_L)/d_\mathrm{NM}= 6.67\times
10^8$ K/m, $\alpha_\mathrm{FI}=0$, and $\Delta T=0$ (blue up-triangles).
Although the thermal gradient in NM ($\alpha_\mathrm{NM}= 6.67\times 10^8$
K/m) is four orders of magnitude larger than that in FI ($\alpha_\mathrm
{FI}=10^4$ K/m), the spin current due to $\alpha_\mathrm{NM}$
(up-triangles) is negligibly smaller than that due to $\alpha_\mathrm{FI}$
(squares), showing ineffective generation of spin current by the thermal
gradient in NM. The purple down-triangles in Fig. \ref{result} are
$j_\mathrm{FI\rightarrow NM}$ for $\alpha_\mathrm{NM}=2.7\times 10^3$ K/m,
$\Delta T=1$ K, and $\alpha_\mathrm{FI}=9.0\times 10^3$ K/m, corresponding to
realistic thermal conductivities of $\kappa_\mathrm{FI}=6.0$
$\mathrm{W/(m\cdot K)}$ and $\kappa_\mathrm{NM}=20$
$\mathrm{W/(m\cdot K)}$ for YIG and Pt \cite{heatpara}, and the
interfacial thermal resistance of $R=1.8\times10^{-5}$ $\mathrm{K/(W/m^2)}$.
Interestingly, the spin current generated by a thermal gradient in FI increase
almost linearly with the temperature while the spin current under a fixed
interfacial temperature difference saturates at a higher enough temperature.

The experimentally measured ISHE voltage $V$ in open-circuit comes from
ISHE-induced charge accumulation. According to above results, when
$T_R>T_L$, spins along the $+z$-direction move to
the $+x$-direction. Due to the ISHE, a charge current flows along the
$+y$-direction (electrons flow to the $-y$-direction), resulting in a
charge accumulation in the front/back surfaces ($xz$-planes in Fig. \ref{system}) and
a higher electric potential in the $+y$ side than that in the $-y$ side
as what was observed in experiment \cite{sse3}. Reversing the direction
of either the magnetization of FI or the temperature gradient, the
ISHE voltage $V$ changes sign. The effective electric field along
the $+y$-direction at the interface can be estimated by \cite{Mosendz}
\begin{equation}
\frac{1}{d_\mathrm{NM}}\int_0^{d_\mathrm{NM}}j_s(x)\theta_\mathrm{SH}\mathrm{d}x=\sigma \mathcal{E}_\mathrm{avg},
\end{equation}
where $\theta_\mathrm{SH}$ is the spin Hall angle \cite{angle} and
$V=\mathcal{E}_\mathrm{avg}w_\mathrm{NM}$, here $w_\mathrm{NM}$
is the width of the NM layer.
Thus, the voltage is given by
\begin{equation}
|V|=\theta_\mathrm{SH}\frac{w_\mathrm{NM}}{d_\mathrm{NM}}\left|\frac{\delta \mu(0)}{e}\right|.
\end{equation}
For $w_\mathrm{NM}=6$ mm (the same as in the experiment \cite{sse3})
and for YIG and Pt parameters, the ISHE voltage is estimated to be
60 $\mathrm{\mu}$V that is larger than the experiment value of 6
$\mathrm{\mu}$V. The agreement is not too bad since a real system
is much more complicated than the ideal model conserded here.
In reality, the thermal parameters and relaxation times
depend on temperature and the structure of a sample.
If these complications can be included, our theory may give a
more accurate estimate of the ISHE voltage for a sample.
In our analysis, we assume the simplest parabolic energy spectrum
and constant relaxation times for magnons and electrons.
Though the physics shall not change, the value of all quantities
should be sensitive to all these parameters. The interface
electron-magnon scattering should be important
for other phenomena in FI/NM structures such as spin pumping
\cite{Kajiwara, Goennenwein,Hillebrands, Wumingzhong}, transverse SSE
\cite{sse1,sse2}, spin transfer torque on FI \cite{SSTMI} and spin Hall
magnetoresistance \cite{NewMR}.
It may also be relevant to the concept of ``spin mixing conductance".

\section{Conclusion}

It is shown that no spin injection and no spin accumulation are
possible at the thermal equilibrium. This conclusion is general
and model-independent as demanded by the thermodynamical laws.
Spin current and spin accumulation can be generated through
electron-magnon scatterings by two thermal sources:
temperature gradients in FI layers and a temperature difference at
NM/FI interface. Both spin current and spin accumulation are sensitive
to material properties. The spin accumulation increases and the
spin current decreases as the spin diffusion length of NM increases.
The spin current arises from imbalance of magnon absorption and
emission originated from different magnon and electron temperatures
or the deviations of magnons from their equilibrium distributions.
Spin current flows from the hotter side to the colder one under a
temperature gradient in FI or under an interfacial temperature
difference, consistent with existing experiments. In contrast,
a temperature gradient in NM cannot efficiently induce a spin current.

\section{Acknowledgment}
This work was supported by National Natural
Science Foundation of China (Grant No. 11374249).
Hong Kong RGC (Grant No. 163011151 and 605413)

\twocolumngrid

\end{document}